\documentclass[sigconf]{acmart}

\renewcommand\footnotetextcopyrightpermission[1]{}
\begin{document}

\title{MDwAIstScheduler: A Low-Cost, Voice-Activated Device for Hands-Free Clinical Scheduling}

\author{Diego Mardian}
\email{dmardia2@asu.edu}
\affiliation{%
  \institution{Arizona State University}
  \city{Tempe}
  \state{Arizona}
  \country{USA}
}

\author{Frank Liu}
\email{fwliu1@asu.edu}
\affiliation{%
  \institution{Arizona State University}
  \city{Tempe}
  \state{Arizona}
  \country{USA}
}

\begin{abstract}
Physicians spend nearly half their workday on EHR tasks and administrative work, contributing to burnout and reducing time for direct patient care. We present MDwAIstScheduler, a low-cost, belt-worn voice assistant that allows hands-free calendar management during patient encounters. Hidden beneath a lab coat, the device avoids the eye-contact disruptions caused by visible screens or wrist-worn devices. Running on a Raspberry Pi with cloud-based speech recognition and LLM intent extraction, the system lets clinicians simply say ``Schedule a follow-up with Mr.\ Smith next Tuesday at 2'' and automatically creates the calendar event. Our demo showcases this end-to-end pipeline.
\end{abstract}

\maketitle

\section{Introduction}
Healthcare providers face an administrative burden that directly impacts their well-being and quality of patient care. Physicians spend only 27\% of their workday in direct clinical time with patients, while 49.2\% is consumed by EHR and desk work~\cite{sinsky2016allocation}. This overload is a significant driver of clinician burnout, which affects over 43\% of physicians in the United States~\cite{shah2025ambient}.

Visible technology also makes the problem worse. Studies show that EHR use during patient encounters reduces physicians' ability to maintain eye contact and actively listen~\cite{alkureishi2016ehr}, and eye-gaze research confirms that physician attention to screens directly influences patient engagement~\cite{asan2014eyegaze}. Wrist-worn devices and desk-mounted systems suffer from the same issue of pulling the physician's gaze away from the patient.

We present \textbf{MDwAIstScheduler}, a belt-worn voice assistant designed to be \textit{invisible} during patient encounters. Hidden beneath a lab coat, the device enables hands-free scheduling without any visual distraction. A physician simply speaks a command ``Schedule a meeting with Dr.\ Patel next Friday'' and the system transcribes, interprets, and creates the calendar event automatically. Unlike commercial ambient AI scribes that focus on documentation and require expensive infrastructure~\cite{tierney2024ambient, ghatnekar2025digitalscribes}, MDwAIstScheduler targets a focused workflow (scheduling) using low-cost hardware, making it practical for smaller practices.

\section{Belt-Worn Wearable Design}

Research on wearable health devices identifies torso-worn devices, including belts, as an emerging category that enables unobtrusive operation without disrupting daily activities~\cite{lu2020wearablehealth, oconnor2021wearable}. For clinician-facing applications, this unobtrusiveness is critical.

A belt-mounted device hidden beneath a lab coat offers key advantages over alternatives:
\begin{itemize}
    \item \textbf{No visual distraction:} The physician never looks at a screen or wrist; the patient sees no device.
    \item \textbf{Preserved eye contact:} Unlike EHR screens that impair communication~\cite{alkureishi2016ehr}, the device is invisible.
    \item \textbf{Hands-free operation:} Voice input requires no manual interaction, keeping hands available for patient care.
    \item \textbf{Natural conversation flow:} Scheduling commands are in normal speech without interrupting the encounter.
\end{itemize}

\begin{figure}[h]
    \centering
    \includegraphics[width=0.32\textwidth]{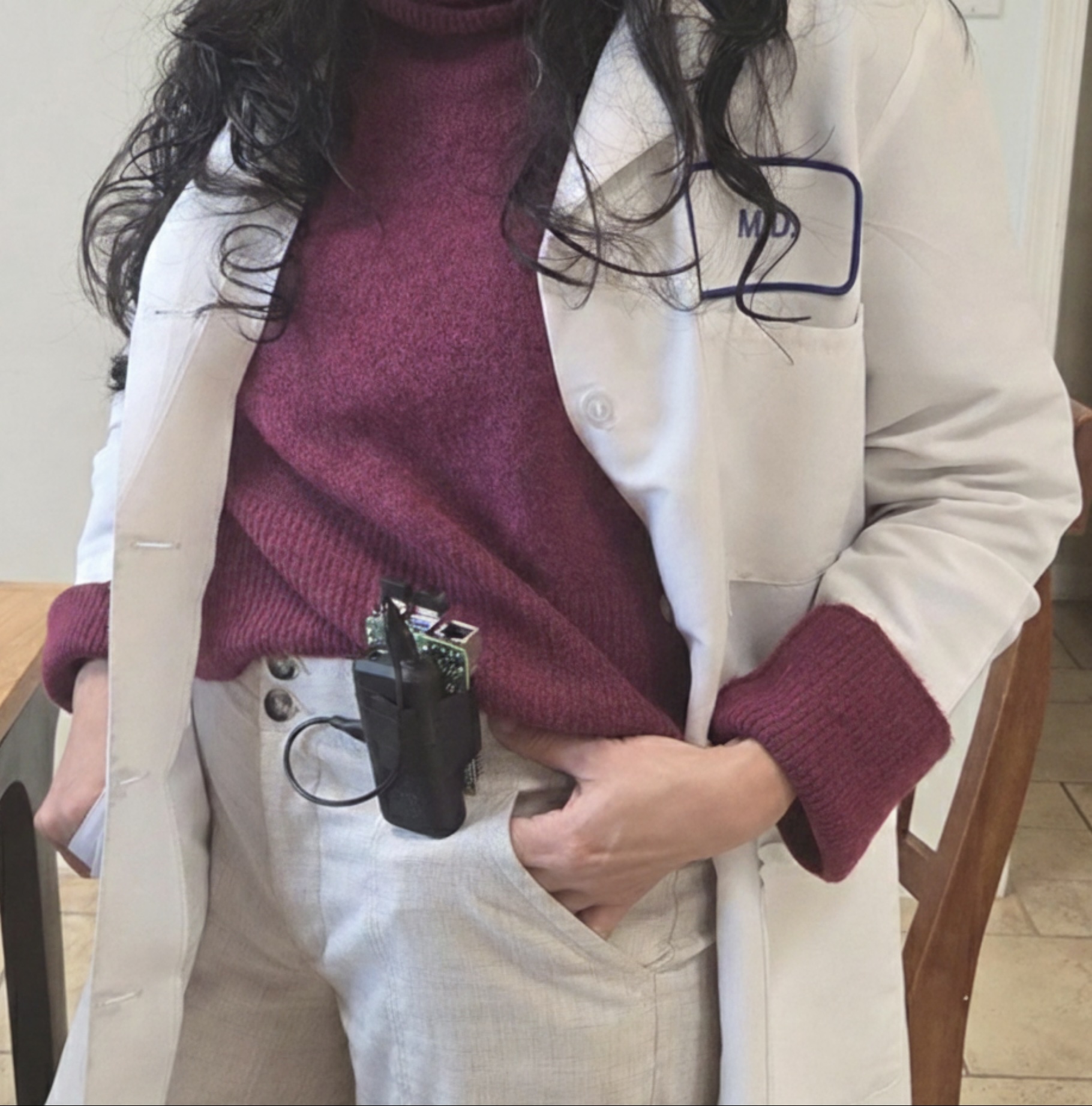}
  \caption{The MDwAIstScheduler system, designed to be worn on the belt and hidden beneath a lab coat during patient encounters.}
    \label{fig:label}
      \centering
\end{figure}

\section{System Architecture}

MDwAIstScheduler runs on a Raspberry Pi 4 (\$35--75) with a USB microphone (\$2--3), keeping total hardware cost under \$100. The system shown in Figure~\ref{fig:label} consists of three components:

\textbf{(1) Continuous Voice Capture.} The device monitors ambient audio and triggers processing when a scheduling command is detected.

\textbf{(2) Speech-to-Text.} Audio is sent to Google Speech Recognition API for transcription. Offloading to the cloud maintains high accuracy on limited hardware, consistent with hybrid edge-cloud architectures~\cite{cuevas2025edgestt, frontiers2025edgellm}.

\textbf{(3) LLM Intent Extraction.} Transcribed text goes to Mixtral-8x7B (via Groq API) with a prompt that extracts scheduling parameters (date, time, duration, attendee, description) as JSON. This replaces brittle rule-based parsing with flexible natural language understanding that handles varied phrasing and implicit references like ``next Monday''~\cite{hodgson2017ehr}. The JSON is then used to create a Google Calendar event via its API.

\section{User Interaction}

Users can (1) speak natural-language scheduling commands to the device, (2) observe real-time transcription and intent extraction displayed on a monitor, and (3) see the resulting calendar event created in Google Calendar. The intermediate JSON output from the LLM to illustrate how natural language is mapped to these structured scheduling parameters is observable as an option. The system can also handle phrasing variations for scheduling.

\section{Discussion and Future Work}

MDwAIstScheduler demonstrates that targeted, low-cost, unobtrusive voice wearables can reduce administrative burden while preserving patient interaction quality. The belt-worn form factor represents a new approach. By remaining hidden, it eliminates the visual distractions of screen-based or wrist-worn alternatives~\cite{alkureishi2016ehr}.

Future work includes: (1) evaluating the system with physicians in family medicine practice to measure impact on time allocation and perceived burden, (2) integrating with EHR scheduling systems beyond Google Calendar, (3) exploring on-device LLM inference for improved privacy~\cite{frontiers2025edgellm}, and (4) adding multi-turn dialogue for disambiguation of ambiguous commands.

\bibliographystyle{ACM-Reference-Format}
\bibliography{references}

\end{document}